# Recent Advances in Metasurface Design and Quantum Optics Applications with Machine Learning, Physics-Informed Neural Networks, and Topology Optimization Methods

Wenye Ji[1, #], Jin Chang[2, *, #], He-Xiu Xu[3, *], Jian Rong Gao[1,4], Simon Gröblacher[2], Paul Urbach[1,*], Aurèle J.L. Adam[1].

1. Department of Imaging Physics, Delft University of Technology, Lorentzweg 1, 2628 CJ, Delft, The Netherlands

2. Department of Quantum Nanoscience, Delft University of Technology, Lorentzweg 1, 2628 CJ, Delft, the Netherlands

3. Shaanxi Key Laboratory of Flexible Electronics (KLoFE), Northwestern Polytechnical University (NPU), 127 West Youyi Road, Xi'an 710072, China

4. SRON Netherlands Institute for Space Research, Niels Bohrweg 4, 2333 CA Leiden, The Netherlands

E-mail: J.Chang-1@tudelft.nl, H.P.Urbach@tudelft.nl, hxxuellen@gmail.com

# These authors contribute equally to this paper.

## Abstract

As a two-dimensional planar material with low depth profile, a metasurface can generate non-classical phase distributions for the transmitted and reflected electromagnetic waves at its interface. Thus, it offers more flexibility to control the wave front. A traditional metasurface design process mainly adopts the forward prediction algorithm, such as Finite Difference Time Domain, combined with manual parameter optimization. However, such methods are time-consuming, and it is difficult to keep the practical meta-atom spectrum being consistent with the ideal one. In addition, since the periodic boundary condition is used in the meta-atom design process, while the aperiodic condition is used in the array simulation, the coupling between neighboring meta-atoms leads to inevitable inaccuracy. In this review, representative intelligent methods for metasurface design are introduced and discussed, including machine learning, physics-information neural network, and topology optimization method. We elaborate on the principle of each approach, analyze their advantages and limitations, and discuss their potential applications. We also summarise recent advances in enabled metasurfaces for quantum optics applications. In short, this paper highlights a promising direction for intelligent metasurface designs and applications for future quantum optics research and serves as an up-to-date reference for researchers in the metasurface and metamaterial fields.

## 1 Introduction

Electromagnetic (EM) wave front modulation has important significance in both scientific researches and industrial applications. It is highly demanded in classical information processing [1], [2], telecommunication [3], [4], military applications [5], [6], imaging systems [7], [8], and their quantum counterparts [9]-[11]. Traditionally, dielectric materials-based devices were used to control EM waves, including lenses and deflectors [12]-[14]. However, conventional dielectric materials have limited choices of dielectric constants. In addition, large dimensions and complex shapes are required to accumulate enough propagating phase to realize targeted functions [15], [16].

In recent years, two-dimensional (2D) planar materials, known as metasurfaces, have solved challenging problems encountered by traditional optical devices. Metasurfaces have artificially tunable EM responses. The modulation of EM waves is achieved by arranging meta-atoms (pixel unit cells of a metasurface) in a predefined order [17]-[38]. The principle is to use the sharp phase variation of the transmitted or reflected wave on the metasurface with the designed structure to effectively control the EM wavefront [39], [40]. After Yu et al. proposed the generalized Snell's law [38], metasurface researches started to flourish and various metasurface devices with different functionalities emerged consistently [41]-[56]. A traditional metasurface design relies on the forward prediction methods: Finite Element Methods or Finite Difference Time Domain Methods are used to predict the optical properties [57]-[58]. Normally, a unit cell is simulated with a periodic boundary. Then more unit cells are combined to form a large area system. Such a process is time-consuming, and the designed meta-atoms are difficult to achieve ideal optical responses. Due to the use of periodic boundary conditions in meta-atom design and aperiodic boundary conditions in the array simulation process, the mutual coupling between different meta-atoms causes unavoidable inaccuracy.

Before we discuss the intelligent optimization techniques for metasurface designs, the traditional methods derived from fundamental laws of physics are introduced in detail as following, which is understandable by logic. The general design process for a metasurface device consists of two steps. The first step is the metasurface unit cell design.

Afterwards, one fills in the discrete metasurface pixels by unit cells with different response including phase, amplitude or other EM properties of a wave [59]. There are several typical mechanisms for metasurface unit cell design. The earliest method is called propagating phase metasurface [60]. The mechanism of propagating phase unit cell is to change EM resonance of the unit cell by adjusting the size of the metal structure that introduces a phase change. This mechanism works on the condition of linear polarization wave. Later on, Pancharatnam-Berry (PB) phase metasurface is proposed [61], which is another mechanism for unit cell design working on circularly polarized waves. The phase change is only related to the rotation angle of unit cell and has thus nothing to do with size. Generally, the additional phase shift of the PB phase unit cell is twice the rotation angle of structure. In design of a PB phase unit cell, according to the PB theory, one generally designs a unit cell structure to achieve phase difference of 180° under the radiation of two orthogonally linear polarization waves. In 2013, Pfeiffer and Grbic proposed Huygens' metasurface to realize perfect transmission with impedance matching theory [62]. The basic theory is as following. One considers the total field on both sides of a metasurface is $\vec{E}$ and $\vec{H}$. Then, equivalent electric current $J_s$ and equivalent magnetic current $M_s$ on the interface of the metasurface are excited. The surface electric impedance $Z_e$ and surface magnetic impedance $Z_m$ are defined by $\vec{E} = Z_e \vec{J}_s$, and $\vec{H} = \frac{1}{Z_m}\vec{M}_s$. According to boundary conditions, reflection and transmission coefficient $R$ and $T$, can be characterized by $Z_e$ and $Z_m$, $R = \frac{-Z_0}{2Z_e + Z_0} + \frac{Z_m}{Z_m + 2Z_0}$, and $T = \frac{2Z_e}{2Z_e + Z_0} - \frac{Z_m}{Z_m + 2Z_0}$, where $Z_0$ is the wave impedance in free space. By derivation, $Z_e$ and $Z_m$ can be characterized by $R$ and $T$, $Z_e = \frac{Z_0}{2}\frac{1+R+T}{1-R-T}$, and $Z_m = 2Z_0\frac{1+R-T}{1-(R-T)}$. We assume the metasurface is lossless. If the transmission amplitude is 1, we derive $R=0$, and $T=e^{i\phi}$. Then, the condition of a high transmission for Huygens' Metasurface is determined by $Z_e = \frac{Z_0^2}{Z_m}$. In other words, as long as one carefully designs $Z_e$ and $Z_m$ of the metasurface, full transmission can be realized [63]. Recently, an interesting metasurface with Anomalous Generalized Brewster Effect (AGBE) is proposed by Fan et al. and Luo et al. [64], [65]. For Brewster Effect, when TM polarized wave (its magnetic field being perpendicular to the plane of incidence) is incident from vacuum to dielectric with permittivity

$\varepsilon_d$, the Brewster's angle $\theta_B$ is $\arctan\sqrt{\varepsilon_d}$ where there is no reflection. However, if a dielectric layer with an appropriate surface impedance $R_s$ is coated on a well-designed metasurface, the reflection-less phenomenon can be realized for different angles, which is called Generalized Brewster Effect (GBE). The relationship between $R_s$ and incident angle $\theta_i$ for TM polarization is determined by $R_s = \dfrac{\sqrt{\varepsilon_d - \sin^2\theta_i}\cos\theta_i}{\sqrt{\varepsilon_d - \sin^2\theta_i} - \varepsilon_d\cos\theta_i}Z_0$, where $Z_0$ is wave impedance of free space. For a refracted wave, refraction angle $\theta_t$ is determined by $\theta_t = \arcsin(\dfrac{\sin\theta_i}{\sqrt{\varepsilon_d}})$. Then, the previous isotropic dielectric is changed into anisotropic dielectric with permittivity $\varepsilon_{pe}=\varepsilon_d$ (perpendicular to refraction direction), and $\varepsilon_{pa}$ (parallel to refraction direction). In this case, the refraction and reflection phenomena remain the same as before. However, if we switch the incident wave from left side of normal ($\theta_i>0$) to right side of normal ($\theta_i<0$), according to the reciprocity principle, the reflection is the same as before, which is determined by $R_s$. While the refraction angle is not only dominated by $\varepsilon_{pe}$, but also $\varepsilon_{pa}$. This interesting phenomenon is called Anomalous Generalized Brewster Effect (AGBE). By carefully designing $R_s$ of a metasurface and permittivity of dielectric, we can control Brewster's and refraction angle. Another metasurface proposed by Chu et al. is the random-flip metasurface [66]. Such a metasurface consists normally of two types of unit cells, which have the same shape and size, but arranged in space inversion positions. The main mechanism behind this metasurface is the reciprocity principle and local space inversion. For demonstration, the Chu et al. designed a metasurface that realizes diffusive reflection while keeping distortion-free transmission. The choice of materials also plays an important role in metasurface designs, which relies on the specific applications and the desired properties of a metasurface, such as operating wavelength, polarization sensitivity, and the level of loss. Generally speaking, metasurfaces are made of subwavelength size metal or dielectric structures. For visible and near-infrared applications, dielectric materials such as Silicon and Titanium Dioxide are often used because of their low loss at those wavelengths compared with metals. For mid- to far-infrared applications, gold and silver are usually used because of their strong resonant property. In microwave applications, metal, such as copper, and high refractive index dielectric materials are normally used because the loss of metal is quite low in microwave band. In other words, the choice of materials for a metasurface depends on different wavelength applications and should keep a balance between material properties such as conductivity, permittivity, and loss.

In a metasurface intelligent design, a commonly used method is the genetic algorithm [51]. However, for multi-parameter problems, solving with the genetic algorithm requires complex computational processes and is time consuming. In addition, with the increased number of parameters, the computational time will increase exponentially. Alternatively, an intelligent metasurface design combining with forward and reverse algorithms offers a solution to overcome the above-mentioned problems encountered by traditional metasurfaces [67], [68]. Compared with traditional optimization algorithms, machine learning can predict unknown problems by learning complex relationships between model variables and optical properties from large known datasets. This strategy can significantly reduce the computational time of a metasurface design by providing a more comprehensive and systematic optimization for metasurface properties [69], [70]. Additionally, by using deep optics and photonics theory, (e.g., Rigorous Coupled Wave Analysis, RCWA) with advanced optimization algorithms (e.g., automatic differentiation, AD), parameters are adjusted and re-evaluated to approach the final goal more efficiently [71], [72]. This method is called topology optimization. We make some comparisons between these methods in terms of physical accuracy, computational time, and degrees of freedom.

In general, to improve the precision of network predictions, machine learning is often applied to simple and fixed metasurface structures, where certain structural parameters are treated as variables [69], [70], [73]-[78]. The precision of the results is directly dependent on the precision of the numerical algorithms used, such as the Finite Element Method (FEM) or Finite-Difference Time-Domain (FDTD), as well as the size of the dataset. Additionally, the inverse design of metasurface structures using machine learning methods may result in non-unique solutions, where the same input may produce different outputs during the training process. To address these limitations, physics-informed neural networks can be employed to accurately predict the EM response of metasurface structures by incorporating physical laws, such as Maxwell's equations and boundary conditions of EM fields, during training [67], [68], [79]-[83]. The network can find the optimal solution by learning the laws of physics. Furthermore, optimization techniques, such as dynamically adjusting the weights, can be used to improve the calculation precision. Topology optimization, the design of the size and shape of the structure within a given space, can also be employed to improve precision [71], [72], [84]-[89]. It considers physical laws, optimization goals, and constraints, such as fabrication precision, and finds a local optimal solution when combined with an optimization algorithm. This method has the highest precision when the physical model is accurate during optimization.

The computational time consumed by machine learning is primarily determined by the size of the data set, the complexity of the metasurface structure, and the number of optimization parameters [69], [70], [73]-[78]. Typically, the computational time can be reduced without compromising the precision of the results, by reducing the size of the required training data set, simplifying the structure, and decreasing the number of optimized parameters. Consequently, the majority of the computational time is concentrated on collecting data sets and training networks. However, this approach may also result in a limited number of design parameters, leading to a reduction in the diversity of structure designs. Physics-informed neural networks, on the other hand, establish a framework based on mathematical and physical methods [67], [68], [79]-[83]. Compared to machine learning methods, they can use fewer data samples to train networks with better generalization capabilities and adjust more structure parameters. This approach not only reduces time loss but also increases the degree of freedom in structure design. The computational time consumed by the topology optimization method is primarily focused on the execution of Generalized Updating Procedures (GUP), which includes optical theory and optimization algorithms, such as gradient descent algorithms [71], [72], [84]-[89]. Therefore, compared to the two previous methods, this one can more quickly approach the optimum. Additionally, this method allows for arbitrary arrangements in space, resulting in the highest degree of freedom in structure design. We give mainly the qualitative comparison here. For further work in this field in the future, specific numerical comparison among these methods needs to be discussed in detail.

The methods discussed above have significantly expanded the possibilities for metasurface designs and have led to notable improvements in performance. A flowchart of the metasurface design process, including principles, fabrication, experimental conditions, and applications, is illustrated in Figure 1. This paper focuses primarily on the recent developments in intelligent metasurface design methods, including machine learning, physics-informed neural networks, and topology optimization. The advantages and limitations of these various methods are analyzed and discussed in detail. We also present recent advances in quantum optics applications enabled metasurfaces before the conclusion section. Thus, this review paper aims to provide a timely summary of recent developments in metasurface design and offers new perspectives for future metasurface designs and applications.

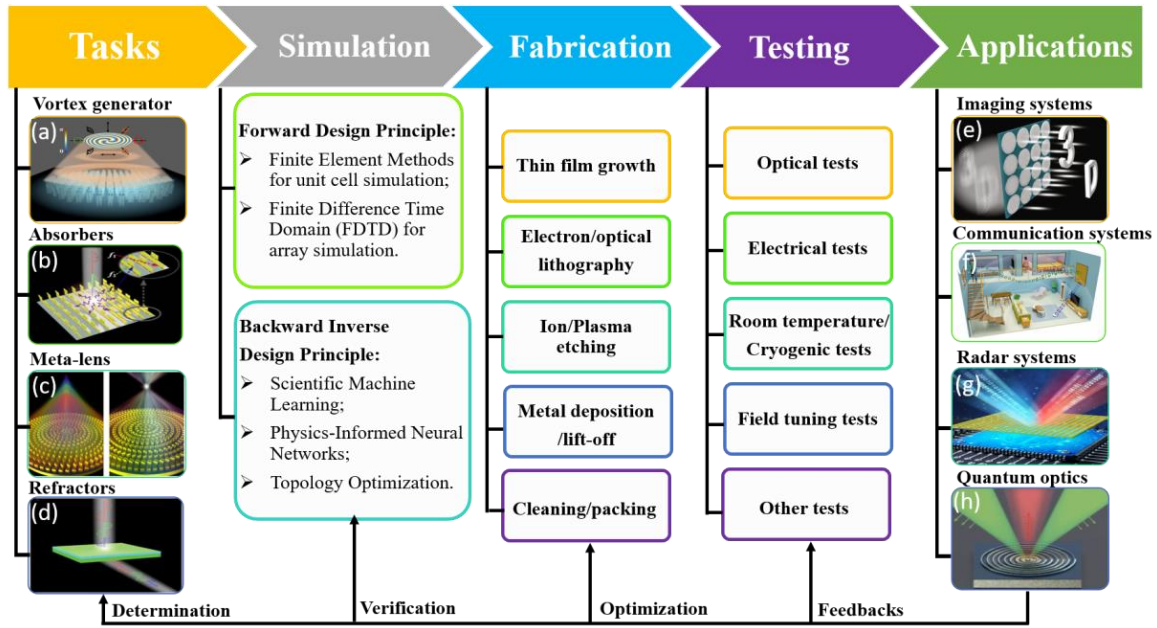

Figure 1. Overview of representative tasks for a metasurface, including (a) vortex generation, reuse with permission for ref. [120], Copyright @ 2022 Song, N. T. et al. (b) light absorption, reuse with permission for ref. [121], © 2019 Optical Society of America under the terms of the OSA Open Access Publishing Agreement, (c) focusing lens, reuse with permission for ref. [122], Copyright @ 2017, Wang, S. M. *et al*. (d) light refraction, reuse with permission for ref. [123], Copyright @ 2021, Ji, W. Y. et al. Laser & Photinics Reviews published by Wiley-VCH GmbH. Besides, the figure also includes metasurface simulation principle, fabrication technics, testing conditions, and typical metasurface applications (e) imaging system, reuse with permission for ref. [7], Copyright @ 2019 Fan, Z. B. *et al*. (f) communication system, reuse with permission for ref. [124], Copyright @ 2020 Zhao, H. T. *et al*. (g) radar system, reuse with permission for ref. [125], Copyright @ 2022 Wan, X. et al. Advanced Intelligent Systems Published by Wiley-VCH GmbH. and (h) quantum optics, reuse with permission for ref. [9], Copyright © 2021, Springer Nature Limited.

## 2 Results (recent advances in metasurface design and analysis)

In this section, we address various approaches and case studies pertaining to the optimization of techniques for designing metasurface-based optical devices. We analyze their underlying principles and methods, and classify them into three main categories, which will be discussed in the subsequent sections.

### 2.1 Machine learning for metasurface design

## 2.1.1 Basic principle

The basic principle flowchart of machine learning approach for metasurface design is illustrated in Figure 2. The general design process is as follows: for a simple cylindrical structure, data sets of EM response can be obtained using forward solver algorithms with a variety of parameter combinations as an input. These data sets can then be used to train a deep neural network, which can calculate the EM response when provided with the input. This is referred to as a forward network. Through the same training process, an inverse network is also obtained. The inverse network differs in that the input is the desired response, and the output is the geometry parameters of the structure. The optimized solutions can also be evaluated using forward solver algorithms to determine if the response is acceptable [73]-[78].

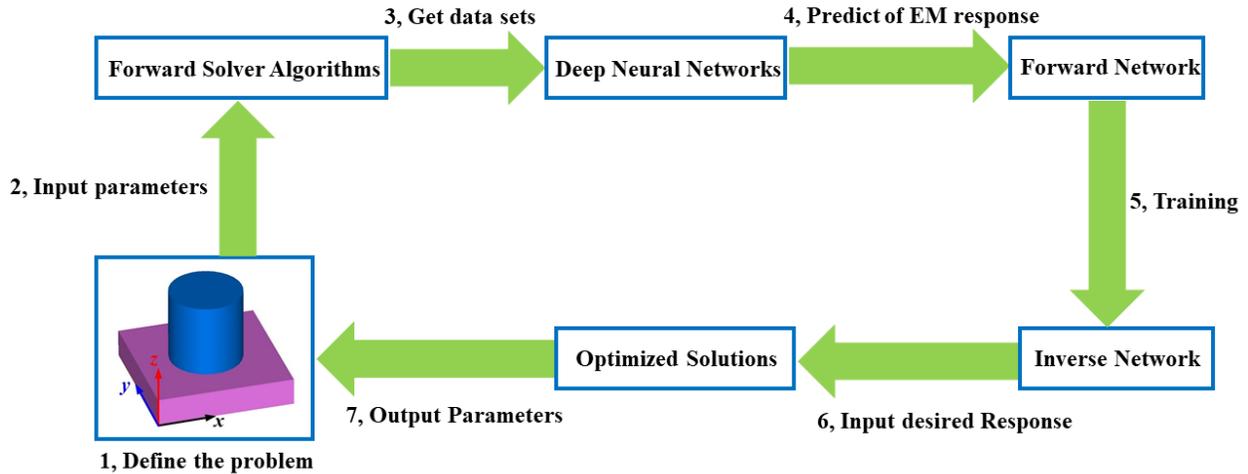

Figure 2 The flowchart steps of basic principle of machine learning for metasurface design: define the problem, collect data sets, pre-process data, train the model, validate the model, train inverse network, input desired response, iterate and optimization, and achieve desired performance.

## 2.1.2 Cases and approaches

In 2019, An et al. proposed a deep learning method for optimizing the optical response of metasurface structures. This method was found to be more accurate and efficient than conventional techniques [70]. Additionally, it represented the first successful application of machine learning to model 3D structures. However, it should be noted that this method is limited to simple, fixed structures, as demonstrated in Figure 3(a). For complex structures, an increase in input size parameters results in a significant reduction of optimization time and accuracy, limiting the degree of freedom. In addition, this work lacks experimental validation of the theoretical and simulation results. Therefore, further researches are needed to establish its validity. In the same year, Zhang et al. proposed a binary coded metasurface structure optimized by machine learning [76], as depicted in Figure 3(b). This structure possesses a higher degree of freedom than the previous work [70], allowing for the use of a large dataset to train the network. Consequently, the network exhibits a high phase response accuracy and rapid optimization speed. The authors [76] also provided experimental validation of the feasibility of their method. However, this work is limited in its focus on coding optimization for unit cell structures and does not provide a design or strategy for array optimization. Furthermore, while the sample works in the microwave band, which allows for easy fabrication of binary coding structures, conversion to the optical band would be hindered by limitations in manufacturing accuracy, making experimental implementation of this method challenging. Later, in 2021, Jiang et al. demonstrated that deep neural networks can predict not only phase, but also group delay for meta-atoms operating in the visible light band [69]. The deep neural network used in this work is illustrated in Figure 3(c). However, if there is a significant deviation in the desired phase spectrum or the spectrum is complex, there will be an inherent error between the network-trained spectrum and the test spectrum, as calculated by the forward solver. In 2022, Ma et al. proposed an innovative approach of designing a multi-functional metasurface by incorporating an optimization algorithm and machine learning in the near-infrared band [77]. The method proposed by Ma et al. allows for multiple functions from a single metasurface to approach the physical limitations. The maxima number of functions that a metasurface can perform is limited by several factors: including physical size, manufacture techniques, coupling between meta-atoms, material loss, and so on. If the area of the metasurface is smaller, the number of resonators which can be designed in one meta-atom period is less, which limits the number of functions. The fabrication techniques used to create resonators also have influence. For example, electron-beam lithography can define highly complex resonator shape, which ensures high precision of the structure. The coupling between neighboring meta-atoms limits the number of the functions because it causes interference between different resonators, which has influence on the functions. Metasurfaces are usually made of metal or dielectric materials with a high refractive index, which can cause large absorption or scattering loss. This factor also influences the functions. The design process is illustrated in Figure 3(d). The authors [77] also conducted experiments and demonstrated their strategy successfully. In the same year, Lin et al. also employed a combination of machine learning and optimization algorithm to optimize all the meta-atoms in a metasurface array in the microwave band [78]. This approach was used to design a retroreflector and a sample was fabricated. It was then effectively demonstrated through an experiment.

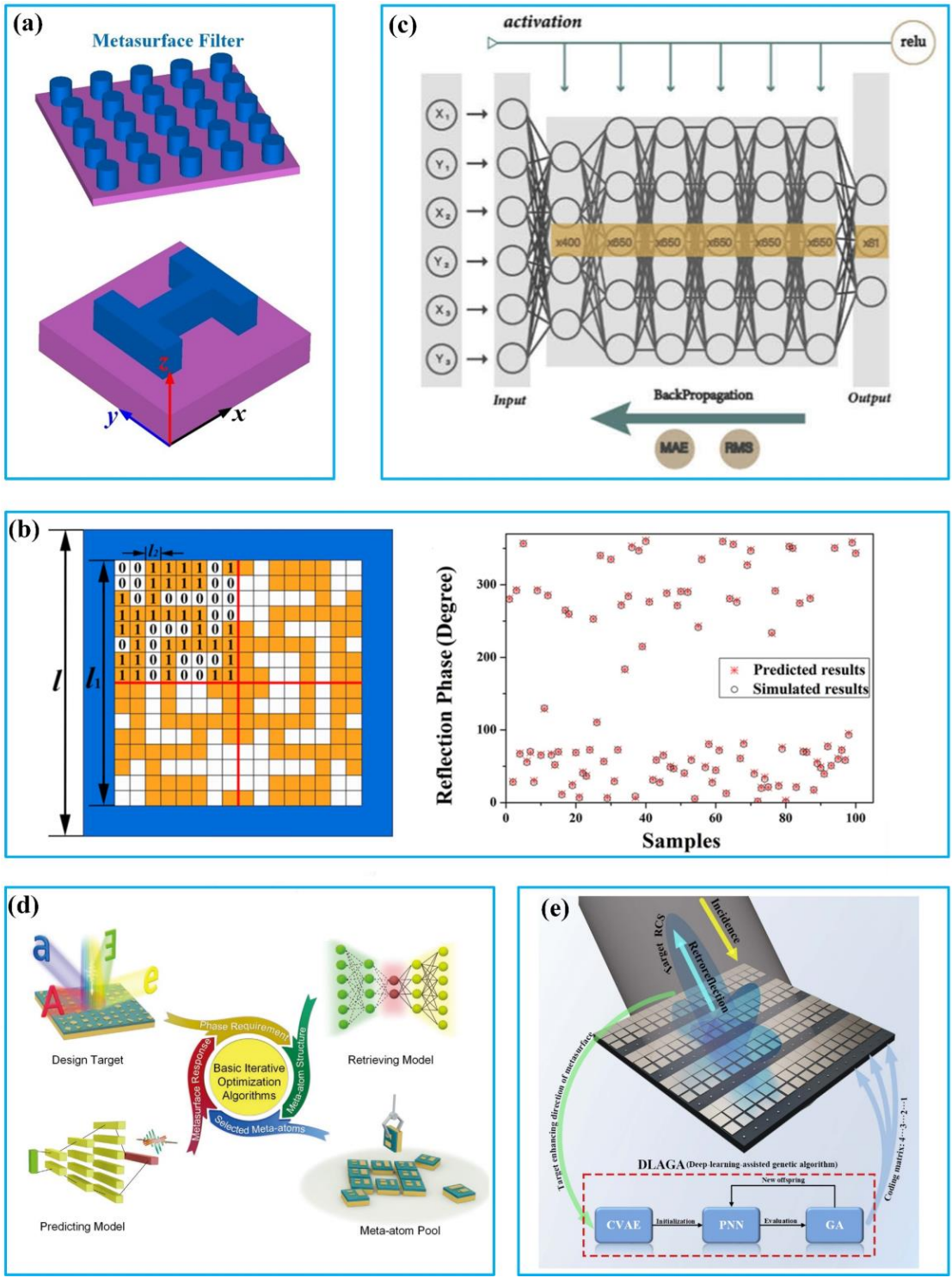

Figure 3. Cases and approaches of machine learning for a metasurface design. (a) Illustration of a metasurface model in the solver. (b) Schematic diagram of binary coded meta-atom and the comparison phase between predicted results by deep learning and simulated results, reuse with permission for ref. [76], © 2018 WILEY-VCH Verlag GmbH & Co. KGaA, Weinheim. (c) Schematic diagram of framework for deep neural network, reuse with permission for ref. [69], © 2021 Optical Society of America under the terms of the OSA Open Access Publishing Agreement. (d) Flowchart of optimization design of EM response by combining machine learning and optimization algorithm, reuse with permission for ref. [77], © 2022 Wiley-VCH GmbH. (e) Flowchart of retroreflector 2-bit array structure design optimized by machine learning, reuse with permission for ref. [78], © 2022 Optical Society of America under the terms of the OSA Open Access Publishing Agreement.

### 2.1.3 Analysis and Conclusion

The strategy to apply machine learning for metasurface design involves the initial creation of a large "library" of potential metasurface designs, followed by the simulation of each design's response using a numerical solver. Subsequently, classical search methods are employed to traverse the library and identify the desired metasurface parameters. This traditional approach is inefficient as well as time-consuming. The machine learning techniques discussed in this section still necessitate the use of a numerical solver to generate training data, but this represents a one-time cost. Once trained, the networks can predict metasurface parameters without the need for further utilization of a numerical solver.

A prevalent issue in the application of general neural network frameworks to metasurface design is the lack of rigor in enforcing the correctness of physics solutions or the manufacturability of discovered parameters. As demonstrated in [69], these frameworks may yield negative parameters, which have no physical significance. This limitation can be addressed through the use of classical constrained optimization frameworks. It is important to eliminate the results that are technically infeasible through validation using numerical solvers. The following methods discussed in this review aim to address this issue.

### 2.2 Physics-informed neural networks for metasurface design

### 2.2.1 Basic principle

The basic principle for physics-informed neural networks is to add information on the physics laws, such as Maxwell's equations or some other partial differential equation (PDE), into neural networks. The operation can be realized by incorporating PDE governing data set into a loss function of framework. Detailed information and theory are referred to the second paragraph of section 2 (Physics-informed neural networks) in [67]. We illustrate a flowchart of physics-informed neural networks for metasurface design in Figure 4. A multi-pillar meta-atom structure, which has a complex design and multiple parameters, results in a greater degree of freedom compared to previous methods. Data sets are obtained by simulating the structure through a forward solver, but with the incorporation of physics laws such as Maxwell's Equations and EM boundary conditions into the neural networks, the required data set size is reduced, leading to a significant reduction in computational time [79]-[83]. The remaining design process is consistent with the machine learning method.

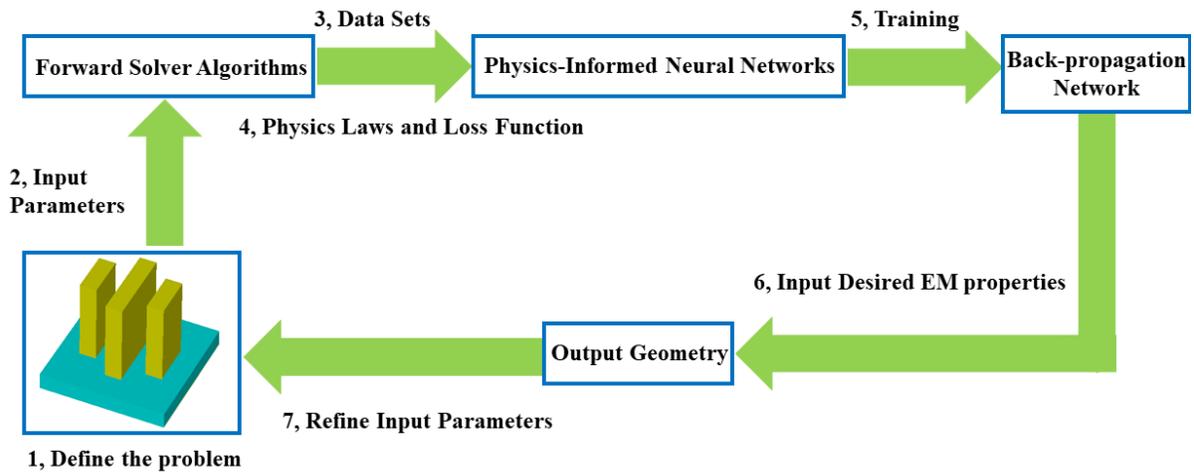

Figure 4 The flowchart steps of basic principle of Physics-Informed Neural Networks for metasurface design: define the problem, create the training data, define physics principle and loss function, define the PINN architecture, validate the network, train backpropagation networks, input desired EM response, iterate the design process, and obtain desired metasurface designs.

### 2.2.2 Cases and approaches

In 2020, Chen et al. utilized a combination of neural networks and partial differential equations, specifically the Helmholtz equation, for metasurface design, the method of which is referred as Physics-Informed Neural Networks [67]. The aim of this work was to use a single cylinder to replace an array of small cylinders, as shown in the first column of Figure 5(a). The goal was to achieve the same electric field response from both objects when a plane wave is incident upon them. As shown in the second column of Figure 5(a), the electric field distribution of the small cylinder array is depicted for a wavelength of 2.1 μm when the wave is incident from left to right. The authors then utilized Physics-Informed Neural Networks to optimize a single cylinder to replace the array. The third column of

Figure 5(a) shows the predicted electric field distribution of the single cylinder, with an error of 2.82% compared to the desired pattern in the second column. This method allows for a reduction in computational time and simplifies the training process, making it useful in the design of invisible cloaks.

In 2022, Tang et al. proposed an approach of incorporating physical guidance and explanation within recurrent neural networks for time response of optical resonances [79]. In this work, A recurrent neural network (RNN) is a kind of neural network (ANN) which is designed to process sequential data, such as time sequence data. RNN processes sequences of inputs and use the information from previous inputs to predict the output. Optical resonance is a phenomenon that occurs when light interacts with a structure, which makes it resonant at a certain frequency. Physics-guided and physics-explainable neural networks are a type of neural network framework that is designed to include physical principles or knowledge into the framework and training process. This method can improve the output accuracy and interpretability because it is consistent with known physics laws or principles. In the first column of Figure 5(b), periodic monolayer graphene stripe structures are expected to produce resonance when a THz wave is incident on it. Then, one can observe time-domain signals. Here, physics-guided and physics-explainable neural networks are adopted to predict the full-length time-domain response, as shown in the second column of Figure 5(b). The input sequence is only 7% of full sequence. With this prediction, they are able to derive the resonant frequency of resonant structures. Additionally, they can obtain more information in the frequency domain by applying the Fourier transform, as depicted in the third column of Figure 5(b). This approach reduces significantly the time required for data collection.

Similarly, Khatib et al. proposed a framework called Deep Lorentz Neural Networks (DLNN) for metamaterials design [68], which is specifically used to model the behavior of EM waves in all-dielectric metamaterials. The DLNN is based on the Lorentz model, which describes how the electric and magnetic properties of a material change in response to an EM field. A schematic of working principle of framework is shown in Figure 5(c). The DLNN is trained by using a dataset from simulated EM wave propagation in all-dielectric metamaterials. Then, it is able to learn physical properties of these materials and predict their behavior according to different inputs. The authors [68] proved that DLNN can accurately predict the behavior of all-dielectric metamaterials with different conditions, such as changing in frequency, polarization, and angle of incidence. By comparing conventional Deep Neural Networks and Lorentz Neural Networks, the authors demonstrate that the latter requires less training data to achieve the same target with less errors. Furthermore, physics-informed neural networks have great universality and generalizability for solving a wide range of optical problems and creating physics models.

Later, Chen et al. updated their physics-informed neural networks with Maxwell equations [83]. It is important to note that the proposed method can predict a 3D distribution of permittivity in an unknown target using near-field data sets. As an example, Figure 5(d) illustrates the cross-sectional plane of the electric field distribution of $E_x$, $E_y$, and $E_z$ (real part) as determined by the finite element simulation method. This information is utilized to train neural networks, which are subsequently combined with wave equations to enable the retrieval of 3D permittivity information from electric field information. This achievement is significant as it allows for the extraction of information from 3D objects, which are more commonly encountered in realistic scenarios. As such, this method has potential applications in fields such as near-field microscopy and medical imaging.

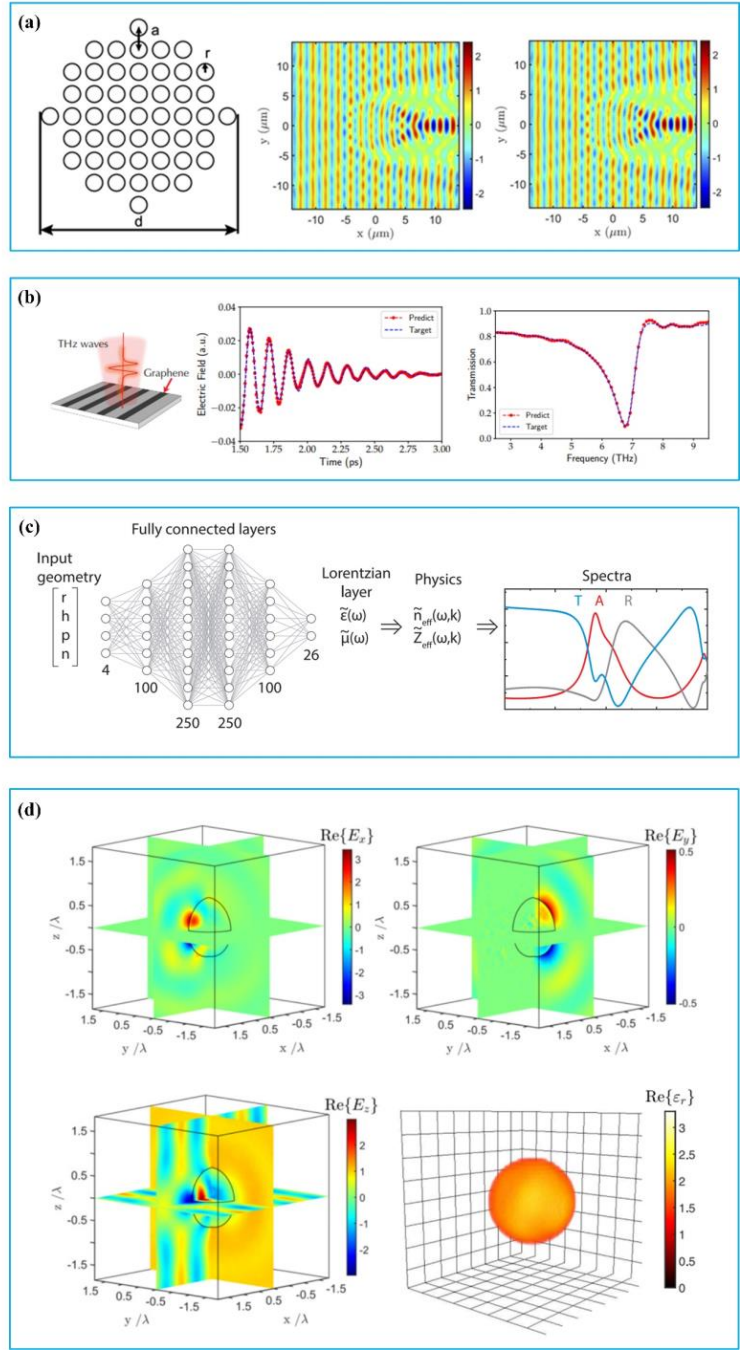

Figure 5. Cases and approaches of Physics-informed neural networks for metasurface design. (a) First column: Schematic diagram of original cylinder array. Second column: Electric field distribution of cylinder array when the plane wave is incident from left to right for the wavelength of 2.1 μm. Third column: Electric distribution of single cylinder predicted by networks, reuse with permission for ref. [67], © 2020 Optical Society of America under the terms of the OSA Open Access Publishing Agreement. (b) First column: Graphene resonant structure. Second column: The comparison of electric field prediction and target signal in time-domain. Third column: The comparison of transmission amplitude prediction and target signal in frequency-domain, reuse with permission for ref. [79], Copyright © 2022, Yingheng Tang et al, under exclusive licence to Springer Nature America, Inc. (c) Schematic diagram Lorentz Neural Networks, reuse with permission for ref. [68], © 2022 Wiley-VCH GmbH. (d) Top left, top right and bottom left: the plane cross sections of electric field distribution of $E_x$, $E_y$, and $E_z$ (real part). Bottom right: 3D permittivity information extracted by physics-informed neural networks, reuse with permission for ref. [83], Copyright © 2022, Chen, Y. Y. This article is distributed under a Creative Commons Attribution (CC BY) license.

### 2.2.3 Analysis and Conclusion

The previous discussions demonstrate that physics-informed neural networks possess a high speed, accuracy, and degree of freedom, and overcome limitations imposed by training data. Additionally, this framework is readily applicable to the problem of recovering geometrical parameters of metasurfaces for a desired EM response. However, it is important to note that a disadvantage of this framework is that a single trained physics-informed neural network cannot be applied to multiple similar inverse problems, requiring retraining for each individual case. Nevertheless, alternative methods that may be equally effective and more efficient without the use of machine learning techniques will be discussed in the following section.

## 2.3 Topology Optimization for metasurface design

### 2.3.1 Basic principle

A flowchart of the topology optimization process for metasurface design is illustrated in Figure 6. The process commences with an initial structure and associated parameters, denoted as $x_i$. The EM response of the structure is then computed using advanced optical theories such as the rigorous coupled-wave analysis (RCWA) method [71], [72]. A loss function, $L$, is subsequently determined by comparing the current EM response to the desired EM response. The gradient of the loss function with respect to the parameters $x_i$ is then determined using gradient algorithms such as automatic differentiation [71]. This gradient information is utilized to update the structure's parameters $x_i$ in the direction that minimizes the loss function. The process is repeated until the loss function reaches its minimum value. The final output is the optimized set of parameters $x_i$ for the desired structure.

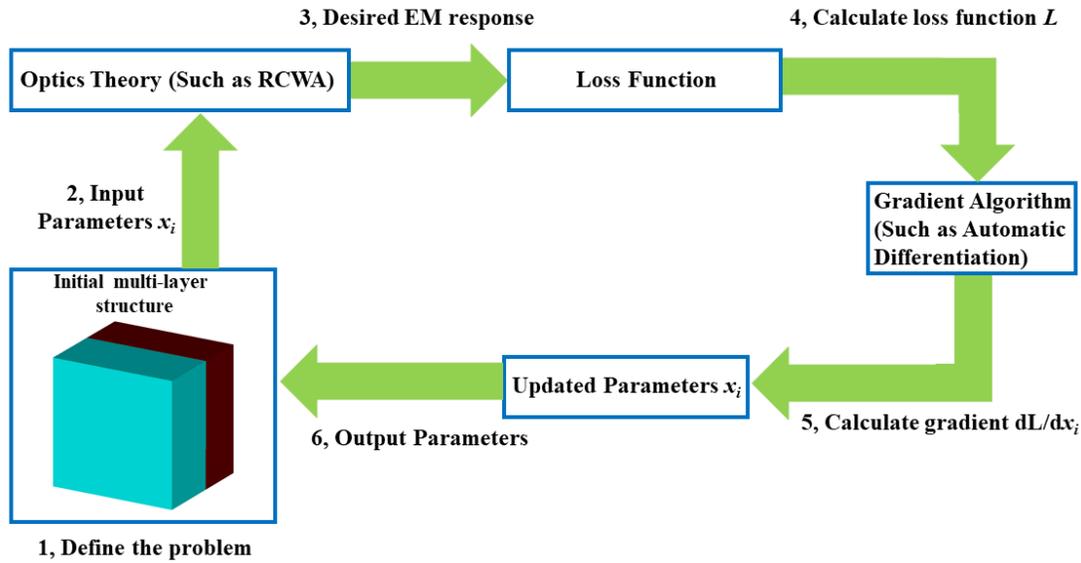

Figure 6 Flowchart steps of basic principle of topology optimization for metasurface design: define the problem, input parameters, make sure desired EM response, calculate loss function, calculate gradient of loss function, update and output parameters, iterate the design process, get desired design.

### 2.3.2 Cases and approaches

In 2019, Lin et al. proposed the use of RCWA method for topology optimization of multi-layer metasurface structures [84]. RCWA is a mathematical approach that is used to analyze and calculate the EM response of multi-layer, periodic structures, making it well-suited for metasurface analysis. Additionally, it has the advantage of highly efficient computation. After obtaining the EM response information of the structure, the authors [88] applied the adjoint method to perform the optimization. As a pure physics-based method, they obtained thousands of degrees of freedom for meta-atoms, which is much more than what achieved using the previous methods. They successfully designed a metalens, as shown in Figure 7(a), but due to the complex structure and high degree of freedom, it is difficult to fabricate such a lens. While the concept is promising, it is currently infeasible to experimentally implement multi-layer, nano-scale structures. In the same year, Phan et al. designed large-area lenses with Aperiodic Fourier Modal Method (AFMM) for finite-sized, isolated devices [85]. Firstly, the authors divided a lens into several small sections for ease of calculation. The curve phase profiles are then linearized for each part, as illustrated in Figure 7(b). For every section, it is composed of aperiodic pillar structures. Importantly, the authors creatively propose an AFMM, which combines a solver for periodic systems and perfectly matched layers. A metasurface with pillar structures can be expressed as a distribution of the permeability and permeability on the surface plane. Then, Stratton-Chu integral equation is used to compute the radiation field in all of space. After optimization, the authors can get the desired field distribution. Also considering the coupling between the edge of neighboring sections, they add a gap of at least 0.2λ between neighboring sections, thereby reducing the coupling between sections. At last, a series of methods adopted above lead to designing a highly efficient lens with high NA. In 2021, Colburn et al. employed the rigorous coupled-wave analysis (RCWA) method in conjunction with automatic differentiation (AD) to optimize metasurface parameters [71]. The overall process is similar to the one that is depicted in Figure 6. AD can be applied to any sequential calculation procedure, but it is complex and involves a series of basic operations. By utilizing the chain rule, the differential coefficient of the complex procedure can be calculated once the differential coefficient for each basic operation is determined. Additionally, the use of parallel calculation on GPUs further increases the speed compared to the adjoint method. An example of designing a lens with several elliptical resonator meta-atoms is shown in the first column of Figure 7(c). As the iteration number increases, the learning curve of focus efficiency improves, as depicted in the second column of Figure 7(c). The normalized electric field intensity in the focal plane is presented in the third column. In the same year, Xu et al. employed RCWA with a multi-object adjoint-based approach to optimize the discrete geometric phase metasurface [89]. Having followed optimization, the discrete structure becomes a continuous structure and the efficiency is significantly improved. The final optimized refractor array and intensity distribution of the electric field are illustrated in Figure 7(d).

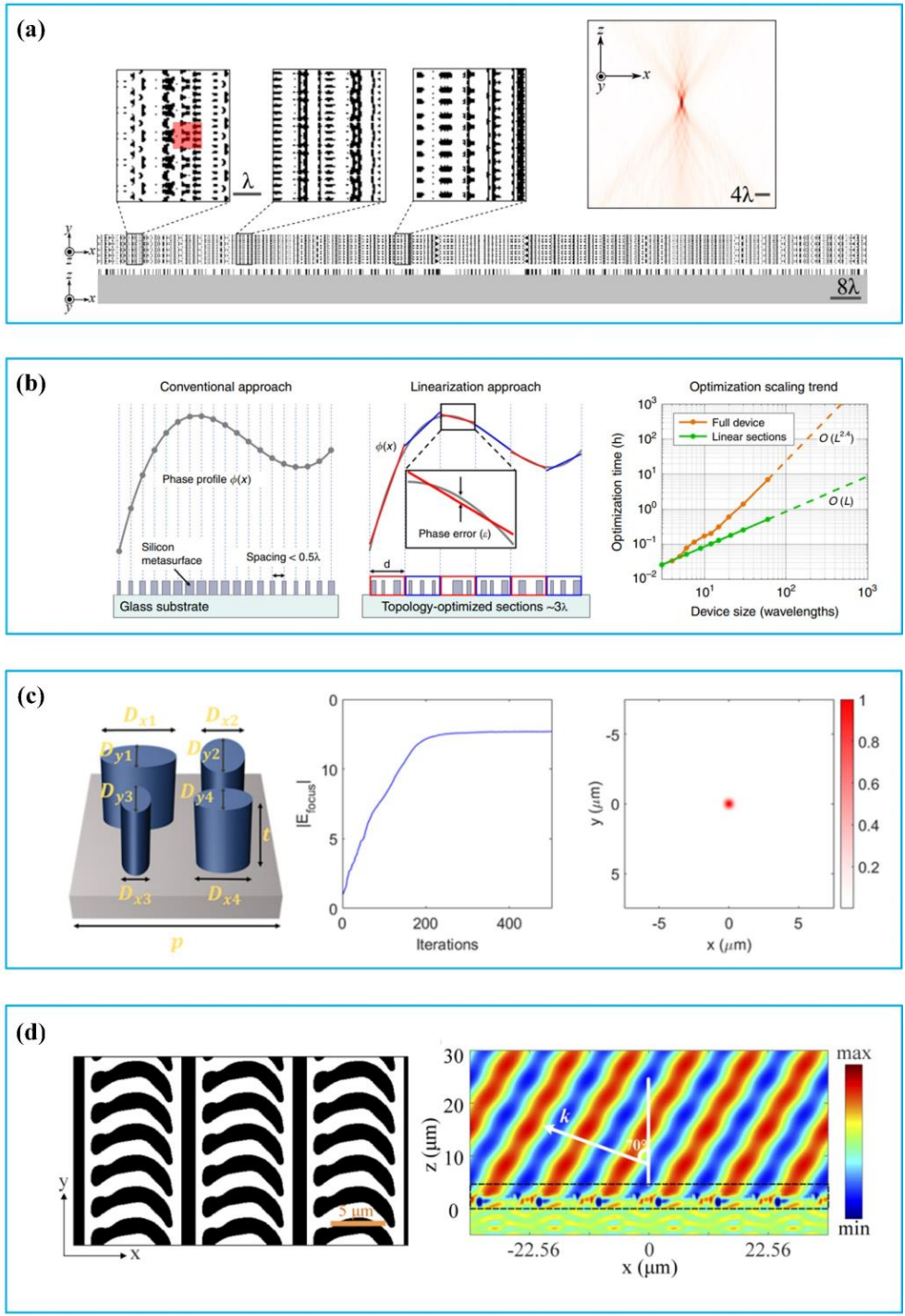

Figure 7. Cases and approaches of topology for metasurface design. (a) Metalens designed by topology optimization method and the focus effect of simulation results for the metalens, reuse with permission for ref. [84], © 2019 Optical Society of America under the terms of the OSA Open Access Publishing Agreement. (b) First column: Conventional lens with curve phase profile. Second column: The topology optimized lens with separated sections and linearization phase profile. Third column: Comparison of optimization time between full device and linear sections varying against device size, reuse with permission for ref. [85], Copyright @ 2019 Phan, T. *et al*. (c) First column: Meta-atom with several elliptical resonators in designing a lens. Second column: The learning curve of focus efficiency varying against iteration. Third column: The normalized electric field intensity in focal plane, reuse with permission for ref. [71], Copyright @ 2021 Colburn, S. *et al*. (d) Left: Topology optimized structure of a metasurface array. Right: Real part of electric field distribution of the refractor, reuse with permission for ref. [89], © 2021 Optical Society of America under the terms of the OSA Open Access Publishing Agreement.

**2.3.3 Analysis and Conclusion**

Based on the previously discussed works, it is evident that in recent years the most effective strategy for metasurface optimization has been through the use of topology optimization methods, which leverage advanced AD formulations or apply high-performance computing techniques. These methods are highly flexible, performant, accurate, fast, and offer a large degree of freedom. The choice of strategy is largely dependent on a specific application. In conclusion, to effectively handle highly complex metasurfaces, a physics-informed neural network can be employed by incorporating elements from these topological optimization approaches, such as the use of the efficient RCWA method, or the implementation of high-performance parallel architecture for training.

**2.4 Metasurfaces for quantum optics applications**

Since the introduction of metasurface concepts, various devices and applications have emerged rapidly, including metasurface antennas [30], [56], radar cross-section modification [29], specialized beam generation [40], [41], active metasurfaces [90], and metasurfaces for quantum optics. In recent years, quantum optics research, as an emerging field, have experienced significant expansion and hold important prospects for applications in areas such as quantum computation, communication, storage, sensing and fundamental quantum physics research [91]-[95]. Metasurfaces are predicted to play a crucial role in the future quantum photonics technology. In this section, we highlight several recent advances in the applications of metasurfaces for quantum optics.

In 2018, Wang et al. proposed the use of flat metasurfaces as a replacement for conventional bulk components to achieve non-classical multiphoton interference [91]. The authors reconstructed one photon and two photon states using a polarization-insensitive detector, as shown in Figure 8(a). The experiment demonstrated the feasibility of controlling multi-photon quantum states using a metasurface. Subsequently, Georgi et al. presented a quantum system that could entangle and disentangle two photon spin states using a metasurface, as shown in Figure 8(b) [92]. The performance of this system was found to be superior to that of conventional optical elements. In 2020, Zhou et al. proposed a polarization-entangled photon source that could function as an optical switch for edge detection mode, as shown in Figure 8(c) [93]. When the photon is in the "switch OFF" or "switch ON" state, the imaging shows a solid or an outlined cat, respectively. In 2022, Gao et al. demonstrated a multi-channel metasurface being capable of transforming polarization-entangled photon pairs [94]. Additionally, it was shown that using two metasurfaces could enable even more channels for entangled photon pair distribution, which holds immense potential for applications in quantum information processing. The intelligent methods discussed in previous sections can improve the performance, accuracy, and speed of a metasurface design, which could potentially be the next-generation design strategy for quantum optics devices.

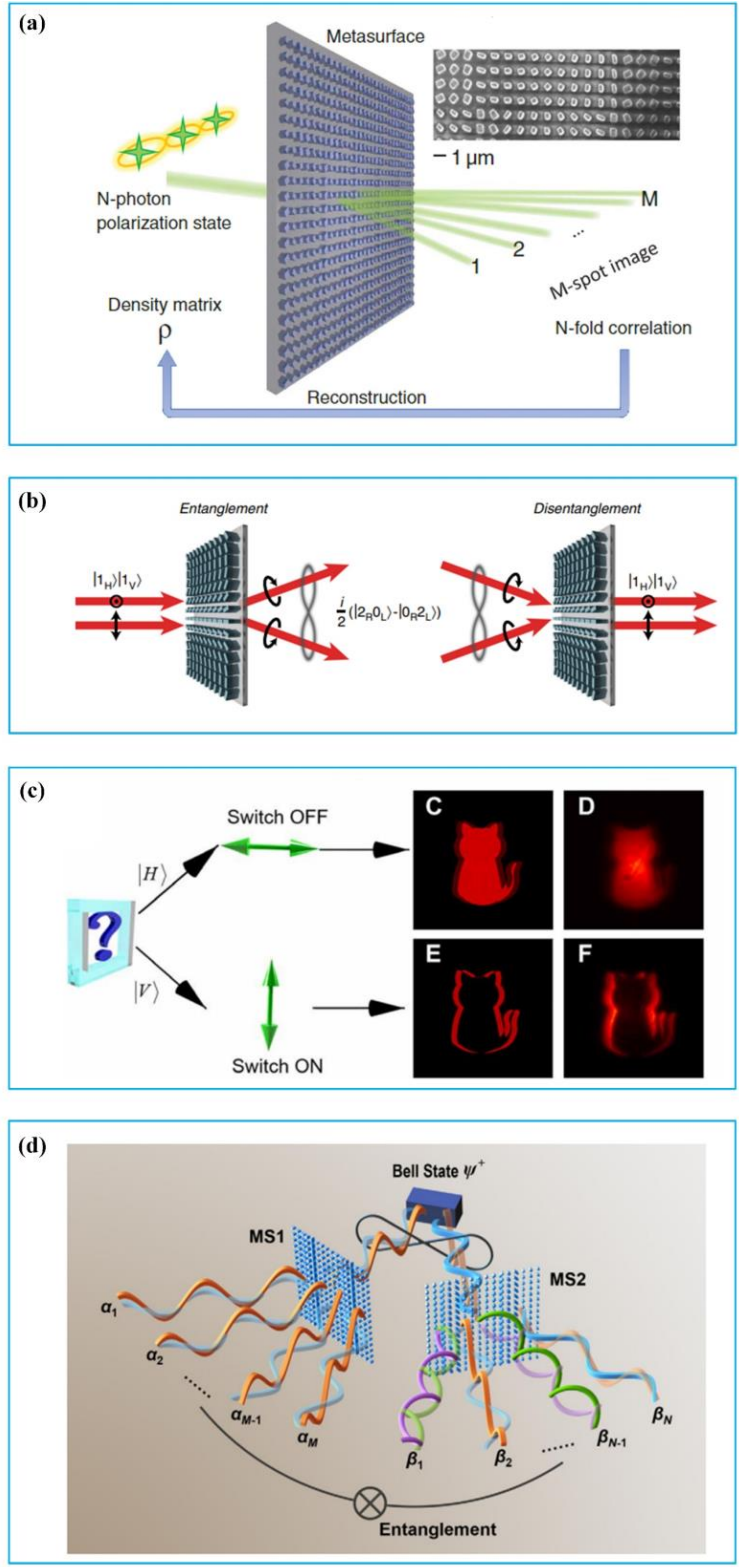

Figure 8. Cases of a metasurface for quantum optics applications. (a) Schematic of quantum state reconstruction with a nanostructured metasurface, reuse with permission for ref. [91], Copyright © 2018 Wang, K. et al. some rights

reserved; exclusive licensee American Association for the Advancement of Science. No claim to original U.S. Government Works. (b) Schematic of entanglement and disentanglement of two photons states with a metasurface, reuse with permission for ref. [92], Copyright @ 2019 Georgi, P. *et al*. (c) Schematic of a metasurface functioning as an optical switch for the optical edge detection mode. When the photon indicates the switch OFF state, the imaging is a solid cat. When the photon indicates the switch ON state, the imaging is an outlined cat, reuse with permission for ref. [93], Copyright © 2020 The Authors, some rights reserved; exclusive licensee American Association for the Advancement of Science. No claim to original U.S. Government Works. Distributed under a Creative Commons Attribution License 4.0 (CC BY). (d) Schematic of two metasurfaces for multi-channel quantum entangle distribution and transformation, reuse with permission for ref. [94], Copyright @ 2022 Gao Y. *et. al.*, PRL, 129, 023601, 2022. American Physical Society, https://doi.org/10.1103/PhysRevLett.129.023601.

## 3 Discussion

In this review, multiple important methods on intelligent design for metasurfaces are elaborated. These methods will become the effective methods for metasurface and metamaterial design in the future and have obvious advantages in physical accuracy and computational time. We give a comparison between forward and backward inverse designs of a metasurface in detail in Table 1. For more examples of using machine learning, physics-informed neural network, and topology optimization method, the readers are referred to [96]-[100]. Furthermore, the methods above can be extended to the design for other optical devices [101], such as photonic crystals [102]-[106], optical cavities [107]-[110], and integrated photonic circuits [111]-[115]. Intelligent metasurfaces are a rapid development direction and have important application prospects in several new revolutionary fields, especially in the field of quantum optics [9]-[11].

Table 1 The comparison between forward and backward inverse design of metasurfacce.

| Design Method | Algorithm | Physical Accuracy | Computational cost |
|---|---|---|---|
| Forward Design Strategy [56]-[58] | Finite Element Method, Finite Difference Time Domain | Only forward solver is difficult to achieve results consistent with ideal optical expectations. | Artificial design and optimization is time-consuming. |
| Machine Learning [69], [70], [73]-[78] | Machine Learning Combined with Numerical Solver Method | Pre-existing numerical forward solver produces physically accurate solutions. | The cost is derived from the training and depends on the training data size, which is one-time cost. |
| Physics-Informed Neural Networks [67], [68], [79]-[83] | Neural Network Combined with Physical Laws behind Physics Process | The existence of solution cannot be presupposed and depends on physical problem. | Less cost, since physical laws restrict the space of admissible solutions to a manageable size. |

| Topology Optimization [71], [72], [84]-[89] | Rigorous Coupled Wave Analysis Combined with Algorithmic Differentiation | The physical predictions of numerical forward solver are accurate, which is the strongest promise of physical accuracy. | The fastest of one iteration of gradient descent is less than others, so the cost is the least. |

Besides academic interests, metasurfaces have emerged as a promising technology with diverse potential applications in industry. Metamaterials have shown remarkable properties such as anomalous reflection, wavefront manipulation, and polarization control. These unique features enable metasurfaces to revolutionize conventional optics and provide innovative solutions for a wide range of industrial applications [116]. In the near future, metasurfaces can be used to enhance the performance of sensors [117], antennas, large sensor arrays, and solar cells, leading to a higher optical efficiency and sensitivity. They can also be used in imaging and holography, providing high-resolution imaging and 3D display capabilities. Furthermore, metasurfaces can be integrated into various devices and systems to improve their functionalities. Therefore, the potential to transform metasurfaces to industrial applications is enormous, and further researches in this field are expected to open new possibilities for their practical implementation [118], [119].

## Conflict of interests

All authors declare no conflict of interests.

## Contributions

W. Ji and J. Chang wrote the manuscript (equal contribution) with the help of H. Xu, J. Gao, S. Gröblacher, P. Urbach and A. Adam.